\documentclass[prd,superscriptaddress,nofootinbib,a4paper,preprintnumbers]{revtex4} 

\usepackage{graphicx}
\usepackage{dcolumn}
\usepackage{bm}
\usepackage{latexsym}
\usepackage{amsfonts}
\usepackage{amssymb}
\usepackage{amsmath}
\usepackage{bbold}
\usepackage{ulem}
\usepackage[utf8]{inputenc} 
\usepackage{xspace}
\newcommand{\Mpl}{{M_{\rm pl}\xspace}}
\begin{document}

\preprint{Imperial/TP/2016/AEG/4, YITP-16-112, IPMU16-0145}

\title{Role of matter in extended quasidilaton massive gravity}

\author{A. Emir G\"umr\"uk\c{c}\"uo\u{g}lu}
\affiliation{Theoretical Physics Group, Blackett Laboratory,
Imperial College London, South Kensington Campus, London SW7 2AZ,
United Kingdom}

\author{Kazuya Koyama}
\affiliation{Institute of Cosmology and Gravitation, University of Portsmouth, Portsmouth PO1 3FX, United Kingdom} 

\author{Shinji Mukohyama}
\affiliation{Center for Gravitational Physics, Yukawa Institute for Theoretical Physics, Kyoto University, 606-8502, Kyoto, Japan}
\affiliation{Kavli Institute for the Physics and Mathematics of the Universe (WPI), The University of Tokyo Institutes for Advanced Study, The University of Tokyo, Kashiwa, Chiba 277-8583, Japan}

\date{\today}
\begin{abstract}
The extended quasidilaton theory is one of the simplest Lorentz-invariant massive gravity theories which can accommodate a stable self-accelerating vacuum solution. In this paper we revisit this theory and study the effect of matter fields. For a matter sector that couples minimally to the physical metric, we find hints of a Jeans type instability in the IR. In the analogue k-essence field set-up, this instability manifests itself as an IR ghost for the scalar field perturbation, 
but this can be interpreted as a classical instability that becomes relevant below some momentum scale in terms of matter density perturbations.
We also consider the effect of the background evolution influenced by matter on the stability of the gravity sector perturbations. In particular, we address the previous claims of ghost instability in the IR around the late time attractor. 
We show that, although the matter-induced modification of the evolution potentially brings tension to the stability conditions, one goes beyond the regime of validity of the effective theory well before the solutions become unstable.
We also draw attention to the fact that the IR stability conditions are also enforced by the existence requirements of consistent background solutions.
\end{abstract}
\maketitle

\section{Introduction}

One of the most intriguing problems in classical field theory has been the construction of a finite range spin--2 theory with local Lorentz-invariance and self interactions. The first linear theory introduced by Pauli and Fierz \cite{bib:Pauli-Fierz}  proved to be incompatible with observations in the small mass limit \cite{bib:vdvz}. Although a non-linear screening mechanism can establish agreement with the predictions of general relativity in this limit \cite{Vainshtein:1972sx}, a generic non-linear completion of Pauli-Fierz theory contains an extra degree of freedom, the Boulware-Deser mode, which violates unitarity \cite{Boulware:1973my}. This issue was resolved only recently by de Rham, Gabadadze and Tolley (dRGT) \cite{deRham:2010kj} who identified the particular tuning which successfully eliminates the dynamics of the Boulware-Deser mode, yielding the long sought for massive spin--2 field theory with five degrees of freedom.

The dRGT theory in the diffeomorphism invariant formulation is constructed out of two metrics: the physical metric $g_{\mu\nu}$ to which the matter couples directly, and the non-dynamical fiducial metric with the restricted form
\begin{equation}
f_{\mu\nu} \equiv \eta_{ab}\partial_\mu \phi^a \partial_\nu\phi^b\,,
\label{eq:fmunudefined}
\end{equation}
where the four scalar fields $\phi^a$ with $a=0,1,2,3$ enjoy Poincar\'e invariance in the field space with metric $\eta_{ab} = {\rm diag}(-1,1,1,1)$. These scalar fields correspond to the gravitational analogue of the St\"uckelberg trick; for a non-trivial field configuration, the fiducial metric corresponds to the flat-space time in a given coordinate system, breaking all four of diffeomorphisms that define general relativity. In particular, in the unitary gauge where $\phi^a= \delta^a_\mu x^\mu$, one has $f_{\mu\nu}= \eta_{\mu\nu}$.  
The graviton mass terms are then constructed by contracting the physical metric and the fixed reference metric through the combination $\sqrt{g^{-1} f}$, where the square-root is shorthand for tensor exponent $1/2$. The dRGT tuning allows only four independent combinations of various powers of $\sqrt{g^{-1} f}$ in the action. 

Being a modified gravity theory in the IR, the dRGT theory attracted considerable attention, particularly in the context of late time cosmology to address the (new) cosmological constant problem with self-accelerating solutions. 
In particular, the theory was found to forbid a Friedmann--Lema\^itre--Robertson--Walker (FLRW) cosmology with a flat spatial geometry \cite{D'Amico:2011jj}, while solutions with negative curvature \cite{Gumrukcuoglu:2011ew} were found to suffer from a non-linear instability \cite{DeFelice:2012mx}. The flat FLRW solutions for the physical metric can be found by allowing an inhomogeneous fiducial metric \cite{inhomogeneous} but these solutions are also plagued by instabilities \cite{instability}.
The only stable cosmological solutions are expected to break homogeneity and/or isotropy (see e.g. \cite{D'Amico:2011jj,bib:anisotropicfriedmann}).

Motivated by the stability of Minkowski solution and the non-existence of a flat FLRW solution, Ref.~\cite{D'Amico:2012zv} introduced an extension of dRGT theory by including an extra scalar field, the ``quasidilaton'', which is associated with the global symmetry
\begin{equation}
\sigma\to\sigma+\sigma_0\,,\qquad
\phi^a \to e^{-\sigma_0/\Mpl}\phi^a\,.
\label{eq:quasidilatonsymmetry}
\end{equation}
Under this transformation, $e^{\sigma/\Mpl}\sqrt{g^{-1} f}$ is invariant and is used to build the mass term for graviton. As a result, the quasidilaton field $\sigma$ acts like the conformal mode of $f_{\mu\nu}$. As the Minkowski background in dRGT is an allowed solution, the conformally flat FLRW background  in the quasidilaton theory is therefore permitted. An interesting property of this cosmology is that it flows to a late time attractor solution 
\cite{Anselmi:2015zva} where the quasidilaton field follows the evolution of e--foldings of expansion. On this attractor, the contribution of the mass term to the total energy density of the universe is effectively a cosmological constant, i.e. it is a self-accelerating solution. Studies of perturbative stability around this background revealed that scalar perturbations are always unstable in the UV \cite{Gumrukcuoglu:2013nza,D'Amico:2013kya}. Although the decoupling limit analysis suggests the existence of stable cosmologies, these correspond to space-times which are homogeneous and isotropic only approximately in the full theory \cite{Gabadadze:2014kaa}.

The instability of FLRW solutions can be avoided by including an additional term to the fiducial metric \eqref{eq:fmunudefined}
\begin{equation}
\tilde{f}_{\mu\nu} \equiv \eta_{ab}\partial_\mu \phi^a \partial_\nu\phi^b - \frac{\alpha_\sigma}{m^2}\,\partial_\mu \left({e^{-\sigma/\Mpl}}\right)\,\partial_\nu
\left({e^{-\sigma/\Mpl}}\right)\,.
\label{eq:extended-fiducial}
\end{equation}
Although the action constructed with the above fiducial metric is known as ``extended quasidilaton theory'' \cite{DeFelice:2013tsa}, the second term above is still allowed by the quasidilatonic transformations \eqref{eq:quasidilatonsymmetry} and preserves the dRGT tuning \cite{Mukohyama:2013raa}.

The FLRW background dynamics in extended quasidilaton theory is dramatically different than the one in the original theory, although at late times, flows to the same late time attractor independently of $\alpha_\sigma$~\footnote{See Ref.~\cite{Kahniashvili:2014wua} for a restricted background analysis, and also the end of Sec.~\ref{sec:kination} for our comment on this work.}. Apart from a finite region of the parameter space, the late time self-accelerating vacuum solutions are stable \cite{DeFelice:2013tsa,DeFelice:2013dua}. 

Given that the cosmological vacuum can be stable, the next step is to investigate the effect of matter fields. In the presence of a matter sector consisting of a single canonical scalar field with a potential, Ref.~\cite{Motohashi:2014una} showed that the stability conditions in vacuum are preserved, although they become time dependent, potentially threatening the stability of a general FLRW background. As an example, the authors considered a scalar field with vanishing potential (occasionally called a ``kination'' field); in this case, the available parameter space for  a stable background at early times becomes of measure zero, leading to an IR ghost in the gravity sector. However, Ref.\cite{Heisenberg:2015voa} argued that 
the stability conditions get modified if the system is sufficiently away from the late time fixed point when the kination field dominates in the early stages.

In this paper we undertake a detailed investigation of the stability of perturbations in the presence of matter. The goal of the paper is twofold: in the presence of a perfect fluid we demonstrate the massive gravity analogue of Jeans instability, which manifests itself as an IR ghost (in the matter sector). We also revisit the purely kinetic scalar field example of Ref.~\cite{Motohashi:2014una} to show that the IR ghost (in the gravity sector) is outside the regime of validity of the effective field theory (EFT). The paper is organized as follows: in Sec.~\ref{sec:theory-and-background} we review the extended quasidilaton theory and discuss the background evolution in the presence of a generic k-essence field. In Sec.~\ref{sec:perturbations}, we introduce cosmological perturbations and obtain the stability conditions. In particular, we show the presence of an IR ghost when the scalar field can be interpreted as an analogue perfect fluid; this ghost instability in the IR is shown to be harmless in Sec.~\ref{sec:canonicaltrans}, where we perform a canonical transformation to change the variable to the density perturbations. In Sec.~\ref{sec:kination}, we consider the case of a canonically normalized scalar field, focusing on the case of a kination field and show that the IR ghost instability associated with it will be outside the reach of the EFT. We conclude with a discussion in Sec.~\ref{sec:discussion}.

\section{Cosmological background}
\label{sec:theory-and-background}
In this Section, we consider the minimal action for the extended quasidilaton field, along with a generic scalar field, minimally coupled to the physical metric, to represent the matter sector. In this construction, we study the evolution of a cosmological background, on the late time de Sitter attractor. 

\subsection{The theory}
We start by reviewing the extended quasidilaton theory. The action in the Einstein frame is given by\footnote{It should be noted that we have omitted two acceptable terms that are allowed by the dRGT tuning, namely the ${\cal L}_0$ term, which corresponds to adding a bare cosmological constant, and the ${\cal L}_1 = [{\cal K}]$ term which is absent to prevent tadpoles. We also remark that one of the four parameters in the mass term $m$, $\alpha_2$, $\alpha_3$ and $\alpha_4$ is redundant. Although it is common to choose $\alpha_2=1$ in the literature, we find that working with three $\alpha_n$ parameters allows us to present our results in an accessible manner.}
\begin{equation}
{S}=\int d^4x \sqrt{-g}\left\{\frac{\Mpl^2}{2} \left[R + 2 m^2 \left(\alpha_2 {\cal L}_2+ \alpha_3 {\cal L}_3+\alpha_4 {\cal L}_4\right)\right]- \frac{\Omega}{2}\partial_\mu\sigma\partial^\mu\sigma + P(X,\chi)\right\}\,,
\label{eq:action}
\end{equation}
where $\Omega$ is a free parameter that controls the strength of the coupling between the quasidilaton and the massive graviton. The graviton mass terms above are tuned in the fashion of dRGT \cite{deRham:2010kj}
\begin{eqnarray}
{\cal L}_{2} & \equiv & \frac{1}{2}\,([{\cal K}]^{2}-[{\cal K}^{2}])\,,\\
{\cal L}_{3} & \equiv & \frac{1}{6}\,([{\cal K}]^{3}-3[{\cal K}][{\cal K}^{2}]+2[{\cal K}^{3}])\,,\\
{\cal L}_{4} & \equiv & \frac{1}{24}\,([{\cal K}]^{4}-6[{\cal K}]^{2}[{\cal K}^{2}]+3[{\cal K}^{2}]^{2}
+8[{\cal K}][{\cal K}^{3}]-6[{\cal K}^{4}])\,,
\end{eqnarray}
which are written in terms of the traces of powers of the building block tensor
\begin{equation}
{\cal K}^{\mu}{}_{\nu}=\delta^{\mu}{}_{\nu}-e^{\sigma/\Mpl}\left(\sqrt{g^{-1}\tilde{f}}\right)_{\ \ \nu}^{\mu}\,,
\end{equation}
where we used the extended fiducial metric $\tilde{f}_{\mu\nu}$ defined in Eq.~\eqref{eq:extended-fiducial}.
The scalar field $\sigma$ is the quasidilaton field associated with the global symmetry in Eq.~\eqref{eq:quasidilatonsymmetry}.
Finally, the matter Lagrangian 
is taken to be an arbitrary function of the scalar field $\chi$ and its canonical kinetic term
\begin{equation}
X \equiv -\frac{1}{2}\partial_\mu\chi\partial^\mu\chi\,.
\end{equation}
This scalar field can be used as a model for an irrotational fluid with the analogue pressure, energy density and sound speed given by \cite{ArmendarizPicon:2000ah}
\begin{equation}
 P = P (X,\chi) \,,\qquad 
\rho \equiv 2P_{,X} X - P, \qquad
 c_s^2 \equiv \frac{P_{,X}}{\rho_{,X}}\,.
\label{eq:fluid}
\end{equation}  

\subsection{Background equations of motion}
\label{sec:bg-general}
We now look for homogeneous, isotropic and spatially flat solutions. In order to preserve these symmetries at the perturbations level, the two background metrics need to respect these in the same coordinate system. To this goal, we take a homogeneous quasidilaton configuration $\sigma= \sigma(t)$ and choose the background St\"uckelberg fields to coincide with the unitary gauge configuration
\begin{equation}
\phi^a = \delta^a_i x^i + \delta^a_0 f(t)\,,
\end{equation}
where for later convenience, $f(t)$ is kept arbitrary to preserve time reparametrization invariance.
In this background, the extended fiducial metric (\ref{eq:extended-fiducial}) becomes
\begin{equation}
ds_{\tilde{f}}^2 = -n(t)^2 dt^2 + \delta_{ij}dx^idx^j\,, \label{eq:extended-fiducial-FLRW}
\end{equation}
where we defined the lapse function through
\begin{equation}
n(t)^2 \equiv \dot{f}^2 +\frac{\alpha_\sigma}{\Mpl^2m^2}\,{\rm e}^{-2 \sigma/\Mpl}\,\dot{\sigma}^2\,.
\label{eq:lapsedef}
\end{equation}

For the dynamical metric $g$, we adopt the flat FLRW ansatz,
\begin{equation}
ds_g^2=-N(t)^2 dt^2 +a(t)^2 \delta_{ij} dx^idx^j\,.
\end{equation}

Finally, we assume a $\chi$ field condensate $\chi=\chi(t)$, with analogue pressure, energy density and sound speed given by \eqref{eq:fluid} on the background $X= \frac{\dot\chi^2}{2\,N^2}$. Bringing all these ansatze together, we find, up to total derivatives, the following mini-superspace action:
\begin{equation}
\frac{S_{\rm mss}}{V} = \Mpl^2\int dt \,a^3 N\,\Bigg\{-\frac{3\,\dot{a}^2}{a^2N^2}-m^2\left[U(\xi)+\frac{r-1}{4}\,\xi\,U'(\xi)\right]+\frac{1}{\Mpl^2} \left[ \frac{\Omega \,\dot{\sigma}^2}{2\,N^2}+P \right]\Bigg\}\,,
\label{eq:minisuperspace}
\end{equation}
where we defined 
\begin{equation}
\xi(t)\equiv \frac{{\rm e}^{\sigma/\Mpl}}{a}\,,\quad
r(t)\equiv \frac{a\,n}{N} = \frac{a\,\sqrt{\dot{f}^2 +\frac{\alpha_\sigma}{\Mpl^2m^2}\,{\rm e}^{-2 \sigma/\Mpl}\,\dot{\sigma}^2}}{N}\,,\quad
U(\xi) \equiv -6 \,\alpha_2 (\xi-1)^2 +4\,\alpha_3 (\xi-1)^3 -\alpha_4 (\xi-1)^4\,.
\label{eq:definitions}
\end{equation}

We now compute the background equations of motion by varying the action (\ref{eq:minisuperspace}) with respect to $N$, $a$, $\sigma$, $\chi$ and $f$. We remark that one of the equations of motion (except $\delta S/\delta N$) can be written as a combination of others through the contracted Bianchi identity
\begin{equation}
\frac{\partial}{\partial t} \frac{\delta S_{\rm mss}}{\delta N} = \sum_{q=a,\sigma,\chi,f} \frac{\dot{q}}{N}\frac{\delta S_{\rm mss}}{\delta q}\,,
\label{eq:bianchi}
\end{equation}
which is satisfied off-shell.

We start with the Friedmann equation, obtained by varying the action with respect to $N$,
\begin{equation}
3\,H^2 = m^2\rho_{m}+\frac{\rho}{\Mpl^2}+\frac{\Omega}{2}\left(H+\frac{\dot\xi}{N\,\xi}\right)^2\,,
\label{eq:EQN}
\end{equation}
where we defined the dimensionless energy density of the mass term as
\begin{equation}
\rho_m \equiv U(\xi)-\frac{\xi}{4}U'(\xi)\,.
\end{equation}
The equation for the acceleration is derived by varying the action \eqref{eq:minisuperspace} with respect to $a$, then using \eqref{eq:EQN}:
\begin{equation}
\frac{2\,\dot H}{N} = m^2J\,\xi\,(r-1)-\frac{\rho+P}{\Mpl^2}-\Omega\,\left(H+\frac{\dot\xi}{N\,\xi}\right)^2\,,
\label{eq:EQA}
\end{equation}
with 
\footnote{In the frequently used 3-parameter setting with $\alpha_2=1$, the three functions $\rho_m$, $J$ and $U'(\xi)$ are no longer independent, satisfying
\begin{equation}
J\Big\vert_{\alpha_2=1} = \xi+\frac{\xi}{(\xi-1)^2}\left(\rho_m+\frac{U'}{4\,\xi}\right)\,.\nonumber
\end{equation}
\label{fn:J--a2=1}}
\begin{equation}
J\equiv \frac{1}{3}\,\frac{\partial}{\partial \xi}\left(U(\xi)-\frac{\xi}{4}U'(\xi)\right)\,.
\end{equation}
The equation of motion for the quasidilaton field $\sigma$ can be written as a second order equation for the ratio $\xi$ defined in Eq.\eqref{eq:definitions}
\begin{equation}
\frac{\Omega}{N\,a^3}\,\frac{d}{dt}\left[a^3\left(H+\frac{\dot\xi}{N\,\xi}\right)\right]
-\frac{\alpha_\sigma}{4\,N\,\xi\,a^4}\,\frac{d}{dt}\left[\frac{a^4 U'(\xi)}{r}\left(H+\frac{\dot\xi}{N\,\xi}\right)\right]=m^2\,\xi\left[3\,J\,(r-1)-U'(\xi)r\right]\,.
\label{eq:EQS}
\end{equation}
The matter field, which is minimally coupled to the physical metric, obeys the usual conservation equation
\begin{equation}
 \frac{\dot\rho}{N}+3\,H\,(\rho+P)=0\,.
\label{eq:EQC}
\end{equation}
Using the definitions (\ref{eq:fluid}) and the above equation of motion, we can also derive the useful relations
\begin{eqnarray}
\dot{P} &=& -3\,N\,c_s^2H\,(\rho+P)+\dot{\chi}(P_{,\chi}-c_s^2\rho_{,\chi})\,,\nonumber\\
\dot{P}_{,\chi} &=& -c_s^2 (\rho_{,\chi}+P_{,\chi})\left(3\,H\,N + \frac{\dot{\chi}\,\rho_{,\chi}}{\rho+P}\right)+\dot{\chi}P_{,\chi\chi}\,,
\label{eq:dotP}
\end{eqnarray}
where a subscript ``$,\chi$'' denotes differentiation with respect to $\chi$ and the time derivative in $\dot{P}_{,\chi}$ always acts after the $\chi$ derivative.

Finally, we calculate the equation of motion for the temporal St\"uckelberg field. Although this equation can be computed through the contracted Bianchi identity \eqref{eq:bianchi}, it is nevertheless useful to obtain it via variation of the action with respect to $f(t)$. Observing that the only $f(t)$ dependence in the action (\ref{eq:minisuperspace}) comes from the function $r(t)$ which contains its first derivative, the St\"uckelberg equation of motion can be readily integrated to give
\begin{equation}
\frac{m^2\Mpl^2\,\dot{f}}{4\,n} \, U'(\xi) \xi= \frac{\kappa}{a^4}\,,
\label{eq:eqstuck0}
\end{equation}
where $\kappa$ is an integration constant whose contribution redshifts as $a^{-4}$ and the quartic polynomial in $\xi$ is given by
\begin{equation}
\xi\,U'(\xi) = 4\,\xi\,(\xi-1)\left[-3\,\alpha_2+3\,\alpha_3(\xi-1)-\alpha_4(\xi-1)^2\right]\,.
\label{eq:eqstuck}
\end{equation}

\subsection{Late time attractor}
\label{sec:bg-fixedpoint}
At late times, as the right hand side of Eq.\eqref{eq:eqstuck0} redshifts away, the system approaches a fixed point solution at $\xi=\xi_{\rm fp}$ given by one of the roots of the quartic polynomial $\xi\,U'(\xi)$. The solution $\xi_{\rm fp}=0$ is unphysical as it leads to strong coupling \cite{D'Amico:2012zv}. The other trivial solution, $\xi_{\rm fp}=1$ does not give rise to self-acceleration so we will also disregard it.\footnote{It should be noted however that $\xi_{\rm fp}=1$ is still a valid fixed point where all five graviton degrees of freedom are dynamical, acquire non-trivial masses and are subject to stability conditions as given in \cite{DeFelice:2013tsa,DeFelice:2013dua}. Although this branch still gives a phenomenology different than general relativity, we will instead concentrate on the fixed points that exhibit self-acceleration.} We are thus left with the remaining two roots of $U'(\xi)$, which are
\begin{equation}
\xi_{\rm fp} =\xi_\pm \equiv 1 + \frac{3\,\alpha_3}{2\,\alpha_4} \pm \sqrt{\frac{9\,\alpha_3^2}{4\,\alpha_4^2}-\frac{3\,\alpha_2}{\alpha_4}}\,.
\label{eq:Xpm}
\end{equation}
The constancy of the ratio $\xi$ on the fixed point completely determines the evolution of the background quasidilaton field $\sigma$ through its definition \eqref{eq:definitions}, giving
\begin{equation}
\frac{\dot{\sigma}}{N\,\Mpl}\Bigg\vert_{\rm fp} = H \left(1+ \frac{\dot{\xi}}{N\,\xi\,H}\right)\Bigg\vert_{\rm fp}= H\,.
\label{eq:sigmafp}
\end{equation}
Thus, in the late asymptotic regime, the contribution to the expansion from the mass term acts as an effective cosmological constant $m^2\rho_m$,
while the contribution of the quasidilaton kinetic energy modifies the effective strength of gravitational interactions. The equations of motion \eqref{eq:EQN}, \eqref{eq:EQA}, \eqref{eq:EQS} on the fixed point attractor thus become
\begin{eqnarray}
\frac{(6-\Omega)}{2}\,H^2 &=&  \frac{\rho}{\Mpl^2}+m^2\rho_m\,,\nonumber\\
\frac{\dot{H}}{N} &=& -\frac{3\,(P+\rho)}{\Mpl^2 (6-\Omega)}\,,\nonumber\\
r &=& 1 + \frac{\Omega}{m^2\,(6-\Omega)\,J\,\xi}\,\left[(6-\Omega)H^2 - \frac{P+\rho}{\Mpl^2}\right]\,,
\label{eq:BG-FIXEDPOINT}
\end{eqnarray}
whereas Eq.\eqref{eq:EQC} continues to hold. 
We stress that the background dynamics on the attractor is independent of the extension parameter $\alpha_\sigma$, so these equations are also valid in the original quasidilaton theory with $\alpha_\sigma=0$ \cite{D'Amico:2012zv}.

\subsection{Allowed parameter space}
\label{sec:parameter}
Let us now discuss the regime of parameters where we can have a sensible evolution. From the Friedmann equation, i.e. the first of \eqref{eq:BG-FIXEDPOINT}, the positivity of the effective gravitational constant imposes 
\begin{equation}
6-\Omega>0\,,
\label{eq:cond1}
\end{equation}
putting an upper bound on the coefficient of the kinetic term of quasidilaton $\Omega$. As we will also show in Sec.\ref{sec:scalars}, this parameter is also bound from below $\Omega>0$ by the stability requirements of the scalar sector \eqref{eq:condUV}.

From here on, we will assume that the effective cosmological constant $m^2\rho_m$ is the main source for the present day accelerated expansion, or less restrictively, that $m^2\rho_m>0$. This leads to the following:
\begin{equation}
(6-\Omega)\,H^2 - 2\,\frac{\rho}{\Mpl^2}>0\,.
\label{eq:cond2}
\end{equation}
In the perturbative analysis, we will encounter terms of the form $(6-\Omega)H^2 - (P+\rho)/\Mpl^2$ whose sign will be crucial in determining the stability. For a fluid with equation of state $w \equiv P/\rho$, satisfying the dominant energy condition $-1<w<1$, we find that (\ref{eq:cond2}) implies
\begin{equation}
(6-\Omega)H^2 - \frac{(1+w)\rho}{\Mpl^2}>0 \,.
\label{eq:cond3}
\end{equation}
For standard constituents of $\Lambda$CDM cosmology, this is a relevant regime. Alternatively, we can express the condition (\ref{eq:cond3}) using the second of Eq.\eqref{eq:BG-FIXEDPOINT}
\begin{equation}
3\,H^2+\frac{\dot{H}}{N} > 0\,.
\label{eq:cond3b}
\end{equation}

We remark that in the presence of a bare cosmological constant, one only needs to shift $U(\xi)\to U(\xi)+\Lambda_{bare}/m^2$, effectively absorbing $\Lambda_{bare}$ into the definition $\rho_m$, and the background equations \eqref{eq:BG-FIXEDPOINT} will still be valid.

From Eq.\eqref{eq:lapsedef}, the positivity of the quantity $(\dot{f}/n)^2$ gives 
\begin{equation}
\frac{\dot{f}^2}{n^2}=1- \frac{\alpha_\sigma H^2}{m^2\xi^2 r^2}\left(1+\frac{\dot{\xi}}{N\,\xi\,H}\right)^2 > 0\,,
\label{eq:fdot}
\end{equation}
which translates into an upper bound on the parameter $\alpha_\sigma$. On the late time attractor, this is simply 
\begin{equation}
\frac{\alpha_\sigma H^2}{m^2\xi^2}< r^2 \,,\quad {\rm for~}\xi=\xi_{\rm fp}\,.
\label{eq:alphabarup}
\end{equation}
As the stability of scalar perturbations in vacuum requires $\alpha_\sigma>0$ \cite{DeFelice:2013tsa} (see also Sec.\ref{sec:scalars}), the ratio $\dot{f}/{n}$ is always constrained to be less than unity from \eqref{eq:fdot}, regardless of the details of the evolution. Along with this information, we see that the only way the past history can accommodate the growth of the right hand side of Eq.\eqref{eq:eqstuck0} is if the ratio satisfies $\xi>1$. This is due to the polynomial dependence on $\xi$ in \eqref{eq:eqstuck0}: since $\xi$ cannot cross the root $\xi =1$ in the past, the fixed point solution will be consistent only if $\xi_{\rm fp}>1$ \cite{Anselmi:2015zva}. Finally, requiring a positive $m^2\rho_m$ and imposing that the tensor graviton mass \eqref{eq:MGW2}, which also corresponds to squared sound speed of vector gravitons \eqref{eq:vectorsoundspeed}, is positive in the late time acceleration domination stage, the allowed parameter regions are:
\begin{eqnarray}
{\cal P}_{++} &:& \Big\{\xi_{\rm fp} = \xi_+
\,,\qquad
\alpha_2m^2>0
\,,\qquad \frac{\alpha_3}{\alpha_2}>0 
\,,\qquad
0<\frac{\alpha_4}{\alpha_2}<\frac{2\,\alpha_3^2}{3\,\alpha_2^2}
\Big\}
\,,\nonumber\\
%
%
{\cal P}_{+-} &:& \Big\{\xi_{\rm fp} = \xi_+
\,,\qquad
\alpha_2m^2<0
\,,\qquad \frac{\alpha_3}{\alpha_2}\leq0 
\,,\qquad
\frac{\alpha_4}{\alpha_2}<0
\Big\}
\,,\nonumber\\
%
%
{\cal P}_{--} &:& \Big\{\xi_{\rm fp} = \xi_-\,,
\qquad
\alpha_2m^2<0
\,,\qquad \frac{\alpha_3}{\alpha_2}>0 
\,,\qquad
0\leq\frac{\alpha_4}{\alpha_2}<\frac{3\,\alpha_3^2}{4\,\alpha_2^2}
\Big\}\,,
\label{eq:allparam}
\end{eqnarray}
where the first subscript of ${\cal P}$ corresponds to the fixed point solution $\xi_\pm$ while the second subscript denotes the sign of the parameter $\alpha_2 m^2$. Notice that there is no allowed region for ${\cal P}_{-+}$.\footnote{The parameter region ${\cal P}_{++}$ is in agreement with those in \cite{Anselmi:2015zva}, where they fixed $\alpha_2=1$ and only considered $m^2>0$.}  It is also worthwhile to mention that in all three allowed parameter regions, the function $J$ satisfies
\begin{equation}
m^2 J >0\,.
\label{eq:Jpositive}
\end{equation}
This condition, combined with \eqref{eq:cond1}, \eqref{eq:cond3} and the third of \eqref{eq:BG-FIXEDPOINT}, gives
\begin{equation}
\frac{r-1}{\Omega}>0\,.
\label{eq:r-1}
\end{equation}
As we will later show in Eq.\eqref{eq:condUV}, the stability of scalar perturbations imposes $\Omega>0$, thus reducing the above inequality to $r>1$.
\section{Cosmological Perturbations}
\label{sec:perturbations}

We now introduce perturbations, by decomposing them based on how they transform under spatial rotations. Thanks to the background symmetry, the tensor, vector and scalar sectors decouple at the linear level. The metric perturbations are given by
\begin{eqnarray}
\delta g_{00} &=& -2\,N^2\,\Phi\,,\nonumber\\
\delta g_{0i} &=& N\,a\,\left(\partial_i B+B_i\right)\,,\nonumber\\
\delta g_{ij} &=& a^2 \left[2\,\delta_{ij}\psi +\left(\partial_i\partial_j-\frac{\delta_{ij}}{3}\partial^k\partial_k\right)E+\partial_{(i}E_{j)}+h_{ij}\right]\,,
\end{eqnarray}
where the latin indices are raised by $\delta^{ij}$ and $\delta^{ij}h_{ij} = \partial^ih_{ij} = \partial^i E_i = \partial^i B_i=0$. The two scalar fields in the system are perturbed as
\begin{equation}
\sigma= \sigma_0 + \Mpl \delta\sigma\,,\qquad
\chi = \chi_0 + \Mpl \delta\chi\,.
\end{equation}
In the remainder of the paper, we omit the subscript $~_0$ of the background quantities for the sake of clarity. We also fix the residual time reparametrization invariance by choosing the physical time coordinate, i.e. $N=1$.

Finally, we keep the St\"uckelberg fields to be purely background, i.e. $\delta \phi^a=0$, thus exhausting the gauge freedom completely. In this set-up, there are in total $12$ degrees of freedom (dof) in the system and no gauge symmetries: $2$ dof in the tensor sector ($h_{ij}$), $4$ dof in the vector sector ($B_i$, $E_i$) and $6$ dof in the scalar sector ($\Phi$, $B$, $\psi$, $E$, $\delta\sigma$, $\delta\chi$). Out of these, $2$ scalar ($\Phi$, $B$) and $2$ vector ($B_i$) dof are non-dynamical. Furthermore, the dRGT tuning allows us to integrate out one more combination. Eventually, we will be left with $2$ tensor dof, $2$ vector dof and $3$ scalar dof. These correspond to the $5$ polarizations of the massive spin--2 field, the quasidilaton and the matter field. We now present the analysis of the perturbations for each sector independently.

\subsection{Tensor modes}
The action quadratic in tensor perturbations is obtained, up to boundary terms, as
\begin{equation}
S^{(2)}_{\rm tensor} = \frac{\Mpl^2}{8}\int d^3k\,dt\,a^3\,\left[\dot{h}_{ij}^\star \dot{h}^{ij}-\left(\frac{k^2}{a^2}+M_{GW}^2\right)h_{ij}^\star h^{ij}\right]\,,
\end{equation}
where the tensor mass is given by
\begin{equation}
M_{GW}^2\equiv\frac{m^2\xi}{\xi-1}\left[[r-2+(2r-1)\xi]J-\frac{(r-1)\xi^2}{(\xi-1)^2}\rho_m\right]\,.
\label{eq:MGW2}
\end{equation}
The stability of the tensor modes is reminiscent of the vacuum case studied in Ref.~\cite{Gumrukcuoglu:2013nza}.
\footnote{The tensor graviton mass found in (\ref{eq:MGW2}) is in agreement with the one given in \cite{Gumrukcuoglu:2013nza} for the vacuum de Sitter solution. In Ref.\cite{Gumrukcuoglu:2013nza}, the parameter choice $\alpha_2=1$ is made, which, using the expression for $J$ on the fixed point from Footnote~\ref{fn:J--a2=1}, implies $\rho_m = (J-\xi)(1-\xi)^2/\xi$, while the vacuum equations of motion with $\dot{H}=0$ give $m^2 J=\Omega H^2/[(r-1)\xi]$.}
They do not exhibit gradient or ghost-like instabilities, although if $M_{GW}^2<0$, it is possible to have a tachyonic instability. On the other hand, just like in the vacuum case, the time-scale of the instability is of the order of $H^{-1}$, so it takes the age of the universe to develop. However, as we will see, the stability of vector modes also relies on the positivity of $M_{GW}^2$, so we will restrict our discussion to the parameter space discussed in Sec.\ref{sec:parameter}.

\subsection{Vector modes}
The action for the vector perturbations is found to be
\begin{equation}
S^{(2)}_{\rm vector} = \frac{\Mpl^2}{16}\int d^3k\,dt\,k^2a^3 \left\{
\dot{E}_i^\star\dot{E}^i-\frac{2}{a}\left(\dot{E}_i^\star B^i+B_i^\star \dot{E}^i\right)-M_{GW}^2 E_i^\star E^i +\frac{4}{a^2}\left[1+ \frac{2\,\Omega\,a^2(3\,H^2+\dot{H})}{3\,k^2(r^2-1)}\right]B_i^\star B^i
\right\}\,.
\end{equation}
Solving for the non-dynamical degree $B_i$, we get
\begin{equation}
B_i = \frac{3\,k^2a\,(r^2-1)}{4\,\Omega\,a^2(3\,H^2+\dot{H})+6\,k^2(r^2-1)}\,\dot{E}_i\,,
\end{equation}
replacing which, the action becomes
\begin{equation}
S^{(2)}_{\rm vector} = \frac{\Mpl^2}{16}\int d^3k\,dt\,k^2a^3 \left[
\left(1+\frac{3\,\tfrac{k^2}{a^2}\,(r^2-1)}{2\,\Omega(3\,H^2+\dot{H})}\right)^{-1}
\dot{E}_i^\star\dot{E}^i-M_{GW}^2 E_i^\star E^i 
\right]\,.
\end{equation}

In order to avoid ghost instability, the following condition needs to be satisfied
\begin{equation}
1+\frac{3\,\tfrac{k^2}{a^2}\,(r^2-1)}{2\,\Omega(3\,H^2+\dot{H})} > 0\,.
\end{equation}
The quantity $(3 H^2+\dot{H})$ is positive for a fluid with $w<1$ \eqref{eq:cond3b}. Thus, in order to have positive kinetic term at all momenta, one needs to impose $(r^2-1)/\Omega>0$. This is precisely the condition we get in the parameter space discussed in Sec.\ref{sec:parameter}, i.e. from Eq.\eqref{eq:r-1}.

We can also calculate the propagation speed of the vector modes by expanding their frequency in the subhorizon limit, $\omega_V^2 = c_V^2 k^2/a^2 + {\cal O}(k^0)$. Requiring that there is no gradient instability imposes
\begin{equation}
c_V^2 = \frac{3\,M_{GW}^2(r^2-1)}{2\,\Omega\,(3\,H^2+\dot{H})} >0\,.
\label{eq:vectorsoundspeed}
\end{equation}
Again, the parameter region \eqref{eq:allparam} already assumes $M_{GW}^2>0$, so the vector modes do not have gradient instability either.

\subsection{Scalar modes}
\label{sec:scalars}
Unfortunately, it is not practical to present the details of the scalar sector calculation, nor it is very informative due to the complicated and opaque expressions. However, we sketch here the steps of the calculation.

Once the action quadratic in scalar perturbations is calculated and plane wave expansion is introduced, there are 6 degrees of freedom, which are $\Phi$, $B$, $\psi$, $E$, $\delta\sigma$ and $\delta\chi$. At the first step, the modes $\Phi$ and $B$ do not have any time derivatives in the quadratic action and their equations of motion give
\begin{eqnarray}
\Phi &=& \frac{3\,c_s^2}{(6-\Omega)(3\,c_s^2 H^2+\dot{H})}\left[
-\frac{\rho_{,\chi}\delta\chi}{\Mpl}+
\frac{\Mpl\,\dot{H}(6-\Omega)}{3\,\dot{\chi}c_s^2}\,\delta\dot{\chi}+\frac{2\,k^2}{a^2}\left(\psi+\frac{k^2}{6}\,E+a\,H\,B\right)
\right.\nonumber\\
&&\left.
\qquad\qquad\qquad\qquad\qquad\qquad\qquad\qquad\qquad\qquad\qquad\qquad- \frac{\Omega(3\,H^2+\dot{H})}{r-1}(\delta\sigma-\psi)-H(\Omega\,\delta\dot{\sigma}-6\,\dot{\psi})\right]\,,
\nonumber\\
B &=& \frac{1}{a\,H(3\,H^2+\dot{H})}\left[[3(r^2-1-\bar{\alpha})H^2-\bar{\alpha}\dot{H}]\delta\sigma-\frac{\Mpl(6-\Omega)(r^2-1)\,H\,\dot{H}}{\Omega\dot{\chi}}\,\delta\chi
\right.\nonumber\\
&&\left.
\qquad\qquad\qquad\qquad\qquad\qquad\qquad\qquad\qquad\qquad\qquad\qquad
+\frac{6(r^2-1)H}{\Omega}\,\left(-H\,\Phi+\dot{\psi}+\frac{k^2}{6}\dot{E}\right)\right]\,,\nonumber\\
\label{eq:solPhiB}
\end{eqnarray}
where
\begin{equation}
\rho_{,\chi} \equiv \frac{\partial}{\partial\chi}\left(2\,P_{,X}X-P\right)=2\,P_{,X\chi}X-P_{,\chi}\,,
\end{equation}
and we defined
\begin{equation}
\bar{\alpha} \equiv \frac{H^2\,\alpha_\sigma}{m^2\,\xi^2}\,.
\end{equation}
Although the above solutions are the most suitable ones for presentation, they are still coupled at this level. By using Eqs.(\ref{eq:solPhiB}) to simultaneously solve for $B$ and $\Phi$, one can obtain an action which depends now on 4 modes, $\psi$, $E$, $\delta\sigma$ and $\delta\chi$. As we mentioned earlier, one combination corresponds to the Boulware-Deser mode, rendered non-dynamical by the dRGT tuning. By redefining the fields through
\begin{equation}
Y_1 \equiv \delta \sigma -\left(\psi+\frac{k^2}{6}\,E\right)\,,\qquad
Y_2 \equiv \delta\chi - \frac{\dot{\chi}}{\Mpl\,H}\,\left(\psi+\frac{k^2}{6}\,E\right)\,,\qquad
Y_3 \equiv \frac{k}{2}\,E\,,
\label{eq:scalarbasis}
\end{equation}
the remaining degree $\psi$ becomes non-dynamical and can be integrated out. Eventually, we obtain an action of the form
\begin{equation}
S^{(2)}_{\rm scalar}= \frac{\Mpl^2}{2}\int d^3k \,dt\,a^3 \,\left(\dot{Y}^\dagger\,K\,\dot{Y} + \dot{Y}^\dagger\,N\,Y- Y^\dagger\,N\,\dot{Y}-Y^\dagger \,M \,Y\right)\,,
\end{equation}
where $K$, $M$ and $N$ are $3\times3$ real, time-dependent matrices with $K^T=K$, $M^T=M$ and $N^T=-N$. 
At this stage the components are not suitable for presentation, although we will present some of the necessary components for the stability discussion.

The signature of the eigenvalues of the kinetic matrix indicates whether the corresponding modes are ghost--like or not. Introducing a rotated basis
\begin{equation}
Z = R^{-1}\,Y\,,
\end{equation}
with the rotation matrix
\begin{equation}
R = \left(
\begin{array}{ccc}
0& 0 & 1\\\\
1 &-\frac{K_{23}}{K_{22}}&\frac{K_{12}K_{33}-K_{13}K_{23}}{K_{23}^2-K_{22}K_{33}} \\\\
 0 & 1 &\frac{K_{13}K_{22}-K_{12}K_{23}}{K_{23}^2-K_{22}K_{33}} 
\end{array}
\right)\,,
\label{eq:basisrotate}
\end{equation}
the kinetic matrix in this new basis becomes diagonal
\begin{equation}
R^T\,K\,R = 
{\rm diag}\left(K_{22}\,,\; K_{33}-\frac{K_{23}^2}{K_{22}}\,,\;\frac{{\rm det}K}{K_{22}K_{33}-K_{23}^2}\right)
\equiv {\rm diag}\left(\kappa_1\,,\;\kappa_2\,,\;\kappa_3\,\right) 
\,.
\label{eq:diagonalkinetic}
\end{equation}
The eigenvalues are as follows:
\begin{eqnarray}
\kappa_1 &=& \frac{\Mpl^2}{\dot{\chi}^2}\,\left[-\frac{3\,c_s^2}{(6-\Omega)\dot{H}}-\frac{\Omega\,(3\,H^2+\dot{H})}{H^2\,\left[\frac{12\,k^2}{a^2}(\bar{\alpha}-1)+\Omega\,(6-\Omega)\,(3\,H^2+\dot{H})\right]}\right]^{-1}\,,
\nonumber\\
\kappa_2 &=& \frac{4\,\Omega\,(3\,H^2+\dot{H})k^2}{r^2\left[\frac{12\,k^2}{a^2}(\bar{\alpha}-1)+ \Omega(6-\Omega)(3\,H^2+\dot{H})\right]}
\left[
\bar{\alpha} + \frac{\Omega(r^2-\bar{\alpha})(6-\Omega)(3\,H^2+\dot{H})}{\frac{12\,k^2}{a^2}\,(r-1)^2}
\right]\,,
\nonumber\\ 
\kappa_3 &=& \Omega +
\frac{\Omega\,(6-\Omega)^2(r^2-\bar{\alpha})(3\,H^2+\dot{H})}{\frac{12\,k^2}{a^2}(r-1)^2}
\,\left[
\bar{\alpha}+\frac{\Omega(r^2-\bar{\alpha})(6-\Omega)(3\,H^2+\dot{H})}{\frac{12\,k^2}{a^2}\,(r-1)^2}
\right]^{-1}\,.
\label{eq:kineticeig}
\end{eqnarray}

Before studying these exact expressions, let us first analyze the subhorizon limit. This will allow us to determine whether the cosmology is UV stable. The kinetic matrix is diagonal at leading order, with $K_{12}={\cal O}(k^{-2})$, $K_{13}=K_{23}={\cal O}(k^{-1})$ and
\begin{equation}
K_{11} = \Omega + {\cal O}(k^{-2})\,,\qquad
K_{22}= \frac{\Mpl^2(6-\Omega)(-\dot{H})}{3\,\dot{\chi}^2c_s^2}+ {\cal O}(k^{-2})\,,\qquad
K_{33}= \frac{\bar{\alpha}\,a^2\Omega \,(3\,H^2+\dot{H})}{3\,r^2(\bar{\alpha}-1)}+ {\cal O}(k^{-2})\,.
\label{eq:subhorizonK}
\end{equation}
The absence of ghosts in the UV, i.e. the positivity of the kinetic terms bring further conditions on the parameters. From background equations of motion \eqref{eq:BG-FIXEDPOINT}, and the positivity of gravitational constant \eqref{eq:cond1}, while assuming dominant energy condition for $\chi$ fluid \eqref{eq:cond3b}, we find that 
\begin{equation}
\Omega > 0 \,,\qquad
\frac{\bar{\alpha}}{\bar{\alpha}-1}>0\,.
\label{eq:condUV}
\end{equation}
The second condition can be satisfied if $\bar\alpha<0$ or $\bar\alpha>1$, and it is strictly non-zero \cite{Gumrukcuoglu:2013nza,D'Amico:2013kya}.

The mixing matrix has only one component at order $k$:
\begin{equation}
N_{12}={\cal O}(k^0)\,,\qquad
N_{13} = \frac{\bar{\alpha}\,\Omega\,(3\,H^2+\dot{H})}{6\,(\bar{\alpha}-1)\,H\,r}\,k+{\cal O}(k^{-1})\,,\qquad
N_{23}={\cal O}(k^{-1})\,.
\end{equation}
Finally, the mass matrix, at order $k^2$, is also diagonal, with $M_{12}={\cal O}(k^0)$, 
$M_{13}=M_{23}=M_{33}={\cal O}(k)$ and 
\begin{equation}
M_{11} = -\frac{\Omega(3\,H^2+\bar{\alpha}\,\dot{H})}{3(\bar{\alpha}-1)a^2\,H^2}\,k^2+{\cal O}(k)\,,\qquad
M_{22} = \frac{\Mpl^2(6-\Omega)(-\dot{H})}{3\,a^2\,\dot{\chi}^2}\,k^2+{\cal O}(k)\,.
\end{equation}

Using these, we can calculate the sound speed of each eigenmode in the subhorizon limit by solving
\begin{equation}
{\rm det} \left[ -K\,\omega^2 -i\,\omega\,(2\,N+\dot{K})+(M+\dot{N})\right] = 0\,,
\end{equation}
where $\omega^2 = \mathbb{1}\left[C_S^2 k^2/a^2 + {\cal O}(k^0)\right]$. Since $\dot{K}$ at order $k$ and $\dot{N}$ at order $k^2$ vanish, they do not contribute and we find the sound speeds for the three dynamical degrees as
\begin{equation}
C_{S,I}^2=1\,,\qquad
C_{S,II}^2=0\,,\qquad
C_{S,III}^2=c_s^2\,.
\label{eq:cs2}
\end{equation}
The first two degrees coincide with the modes in the vacuum case \cite{DeFelice:2013tsa} while the last mode clearly corresponds to the matter perturbation.
Thus we have established that the stability conditions for the scalar sector in the UV are same as in the vacuum case, with the additional requirement that the equation of state for the matter field satisfies $-1<w<1$.

We now turn to the small momentum limit, $k\ll a\,H$. From Eq.(\ref{eq:kineticeig}), we find
\begin{eqnarray}
\kappa_1 &=& \frac{\Mpl^2H^2(6-\Omega)(-\dot{H})}{(3\,c_s^2H^2+\dot{H})\dot{\chi}^2}+{\cal O}(k^2)
\,,
\nonumber\\
\kappa_2 &=& \frac{\Omega\,a^2(r^2-\bar{\alpha})(3\,H^2+\dot{H})}{3\,(r-1)^2r^2}+{\cal O}(k^2)\,,
\nonumber\\
\kappa_3 &=&6+{\cal O}(k^2)\,.
\label{eq:condIR}
\end{eqnarray}
The requirement that the last two terms are positive, along with the conditions (\ref{eq:condUV}), give
\begin{equation}
0<\Omega<6\,,\qquad 1<\bar{\alpha}<r^2\,, 
\label{eq:conditions}
\end{equation}
where we discarded the option $\bar{\alpha}<0$, which satisfies both IR and UV no-ghost conditions, but makes the kinetic terms $\kappa_2$ and/or $\kappa_3$ negative at intermediate momenta. 
It can be verified from Eq.(\ref{eq:kineticeig}) that for an expansion satisfying $\dot{H}<0$ and $3\,H^2+\dot{H}>0$, the conditions (\ref{eq:conditions}) are sufficient to make the kinetic terms $\kappa_2$ and $\kappa_3$ positive at any momenta. We also remark that the upper bounds for $\bar{\alpha}$ and $\Omega$ coincide with the ones imposed by the existence requirement of background solution \eqref{eq:alphabarup} and positive gravitational constant \eqref{eq:cond1}, respectively.

On the other hand, the first kinetic eigenvalue, which corresponds to matter field perturbations, can be positive at arbitrarily low momenta only if
\begin{equation}
3 + \frac{\dot{H}}{c_s^2H^2}>0\,.
\label{eq:condnontrivial}
\end{equation}
An alternative approach to obtain this condition is to use the exact expressions (\ref{eq:kineticeig}). Then, it is evident that there is a critical momentum
\begin{equation}
\frac{k_c}{a} = \sqrt{-\frac{\Omega(6-\Omega)(3\,H^2+\dot{H})\left(3+\frac{\dot{H}}{c_s^2\,H^2}\right)}{36(\bar{\alpha}-1)}}\,,
\label{eq:criticalk}
\end{equation}
below which, the second (negative) term in the denominator of $\kappa_1$ dominates. When $k=k_c$, the kinetic term $\kappa_1$ diverges; if the mass matrix and non-linear interactions are finite at this point, the degree of freedom corresponding to $\kappa_1$ is weakly coupled. In other words, the transition from stable to ghost degree can proceed dynamically within the regime of validity of the EFT. To impose stability of all modes with arbitrary $k$, the critical momentum needs to be imaginary, i.e. the condition (\ref{eq:condnontrivial}) should be satisfied.

For a scalar field with a canonical kinetic term and a field dependent potential, we have $P(X,\chi)=X-V(\chi)$ with unit sound speed, so the condition \eqref{eq:condnontrivial} coincides with the weak energy condition \eqref{eq:cond3b}. In this very simple scalar field theory, the field perturbation is not a ghost under already assumed conditions. This case is discussed in detail in Sec.\ref{sec:kination}.

However, it is straightforward to devise a case where \eqref{eq:condnontrivial} no longer holds. If we consider the $\chi$ fluid to be non-relativistic matter with an effective equation of state $w=0$, then during the matter dominated stage we have $\dot{H}=-3\,H^2/2$. In order to have no ghost degrees at any momenta, the condition reduces to $c_s^2>1/2$, which is clearly relativistic. 

More generally, let us consider a perfect fluid with constant equation of state, and assume $\chi\to\chi+\chi_0$ invariance in the action. In this case, we have $c_s^2=w$, and the expression for the critical momentum \eqref{eq:criticalk} can be written as
\begin{equation}
\frac{k_c}{a\,H} = \frac{\vert 1-w\vert}{4} \sqrt{\frac{\Omega(6-\Omega)}{w\,(\bar\alpha-1)}}\,,
\end{equation}
which can be imaginary only if 
\begin{equation}
w<0\,.
\end{equation}
However, this choice renders the second kinetic term \eqref{eq:subhorizonK} negative in the UV for the perfect fluid limit $c_s^2=w$. 
Finally, for a dust fluid with $w=0$, the critical momentum diverges to infinity, so na\"ively matter perturbations are ghost--like at any momenta. 
 
This is an IR instability of the matter which exhibits itself as a ghost instability. As we argue in the next Section, its presence can be traced back to the choice of the field perturbation as the fundamental variable. We note that this instability is remarkably different than the IR instability found in \cite{Motohashi:2014una}. Specifically, the instability in this Section turns up in the matter sector, when $1/\kappa_1$ crosses zero. In contrast, the instability in \cite{Motohashi:2014una} emerges in the gravity sector, when $\kappa_2$ crosses zero. The latter case will be discussed in detail in Sec.\ref{sec:kination}.

\section{Exorcising the ghosty matter perturbations}
\label{sec:canonicaltrans}
The matter perturbations becoming ghostlike in the IR does not necessarily imply a catastrophic instability of the cosmological model \cite{Gumrukcuoglu:2016jbh}. Instead, it may be an indication of a physical feature associated with a mild instability. To clarify this point, we now perform a canonical transformation and treat the energy density perturbation of the matter field as the fundamental variable. In this picture it becomes clear that the IR ghost instability that we encountered in the perfect fluid analogue is actually a classical instability that becomes relevant below some momentum scale, much like the Jeans instability. 

Instead of using the scalar field perturbation as the fundamental variable, we therefore choose energy density perturbations. For the scalar field with Lagrangian $P(X,\chi)$, the matter energy density is given by the second of Eq.\eqref{eq:fluid}, which at first order in perturbations reads
\begin{equation}
\bar{\delta \rho} = \frac{\rho+P}{c_s^2}\left(\Mpl\,\frac{\delta\dot\chi}{\dot\chi}-\Phi\right) + \Mpl\,\delta\chi\left(P_{,X\chi}\dot{\chi}^2-P_{,\chi}\right)\,.
\label{eq:drhodefined}
\end{equation}
Inspecting the non-reduced action quadratic in scalar perturbations, we find that the only reference to the $|\delta\dot\chi|^2$ appears in the combination $c_s^2 \bar{\delta\rho}^2/(2(\rho+P))$. With this information, we introduce an auxiliary field $\delta\rho$ to the action as follows:
\begin{equation}
\tilde{S}^{(2)}_{\rm scalar} = S^{(2)}_{\rm scalar} -\int d^3k\,dt\,a^3 \frac{c_s^2}{2\,(\rho+P)}\,\vert \delta\rho-\bar{\delta\rho}\vert^2\,,
\label{eq:newscalaraction}
\end{equation}
where $\bar{\delta\rho}$ is the expression given in \eqref{eq:drhodefined} in terms of the scalar field perturbations, while $\delta\rho$ is the newly introduced auxiliary variable. The coefficient of the second term is chosen such that the coefficient of the $|\delta\dot{\chi}|^2$ term in
$\tilde{S}^{(2)}_{\rm scalar}$ vanishes. Notice that varying the above action wrt $\delta\rho^\star$, we find
\begin{equation}
\delta\rho =\bar{\delta\rho}\,,
\end{equation}
and the two actions $\tilde{S}^{(2)}_{\rm scalar}$ and $S^{(2)}_{\rm scalar}$ coincide. However, we instead vary the action \eqref{eq:newscalaraction} with respect to the scalar field perturbation $\delta\chi$, whose time derivatives can now be removed by adding boundary terms. Its equation of motion gives
\begin{equation}
\delta\chi = -\frac{\dot\chi}{\Mpl(\rho+P){\cal Q}^2}\left[\delta\dot\rho+3(1+c_s^2)H\,\delta\rho+(\rho+P)\left(\frac{k^2B}{a}+3\,\dot{\psi}\right)\right]\,,
\label{eq:soldchi}
\end{equation}
where we defined
\begin{equation}
{\cal Q}^2 \equiv 
\frac{k^2}{a^2}+9\,H^2 \left(c_s^2-\frac{\dot{P}}{\dot{\rho}}\right)
=\frac{k^2}{a^2}+
\frac{3\,H\,\dot\chi}{\rho+P} \Bigg((1+c_s^2)P_{,\chi}-c_s^2P_{,X\chi}\dot\chi^2\Bigg)\,.
\label{eq:Q2def}
\end{equation}
We note that for a shift symmetric matter action, invariant under $\chi\to\chi+\chi_0$, the second term in parentheses vanishes and the quantity ${\cal Q}$ coincides with the physical momentum.

Using the solution \eqref{eq:soldchi} back in the action, we recover the same number of variables as the starting point, removing $\delta\chi$ in favor of $\delta \rho$. This procedure is equivalent to a canonical transformation \cite{DeFelice:2015moy}. From this point on, the calculation proceeds as in Sec.\ref{sec:scalars}. We first integrate out the modes $\Phi$ and $B$ whose equations of motion now read:
\begin{eqnarray}
\Phi &=& \frac{1}{(6-\Omega)H^2}\left[
-\frac{\delta\rho}{\Mpl^2}+
\frac{2\,k^2}{a^2}\left(\psi+\frac{k^2}{6}\,E+a\,H\,B\right)
- \frac{\Omega(3\,H^2+\dot{H})}{r-1}(\delta\sigma-\psi)-H(\Omega\,\delta\dot{\sigma}-6\,\dot{\psi})\right]\,,
\nonumber\\
B &=& \frac{\Omega\,a\,{\cal Q}^2}{H\,\left[
\Omega\,a^2{\cal Q}^2(3\,H^2+\dot{H})-
k^2(6-\Omega)(r^2-1)\dot{H}\right]}
\nonumber\\
&&\qquad\times
\left[[3(r^2-1-\bar{\alpha})H^2-\bar{\alpha}\dot{H}]\delta\sigma
-\frac{3(r^2-1)H}{\Omega\,{\cal Q}^2\Mpl^2}
\left[\delta\dot\rho+3(1+c_s^2)H\,\delta\rho+3\,(\rho+P)\dot{\psi}\right]
\right.\nonumber\\
&&\left.
\qquad\qquad\qquad\qquad\qquad\qquad\qquad\qquad\qquad\qquad\qquad\qquad\qquad
+\frac{6(r^2-1)H}{\Omega}\,\left(-H\,\Phi+\dot{\psi}+\frac{k^2}{6}\dot{E}\right)\right]\,.
\label{eq:solPhiBcanonical}
\end{eqnarray}
After using the solutions for $\Phi$ and $B$ in the action, we perform a linear transformation on the field basis and define:
\begin{equation}
\tilde{Y}_1 \equiv \delta \sigma -\left(\psi+\frac{k^2}{6}\,E\right)\,,\qquad
\tilde{Y}_2 \equiv \delta\rho - \frac{\dot{\rho}}{H}\,\left(\psi+\frac{k^2}{6}\,E\right)\,,\qquad
\tilde{Y}_3 \equiv \frac{k}{2}\,E\,,
\label{eq:scalarbasiscanonical}
\end{equation}
and as in Sec.\ref{sec:scalars}, the field $\psi$ becomes non-dynamical as a consequence of dRGT tuning. Once this field is integrated out, the action becomes formally
\begin{equation}
\tilde{S}^{(2)}_{\rm scalar}= \frac{\Mpl^2}{2}\int d^3k \,dt\,a^3 \,\left(\dot{\tilde{Y}}^\dagger\,\tilde{K}\,\dot{\tilde{Y}} + \dot{\tilde{Y}}^\dagger\,\tilde{N}\,\tilde{Y}- \tilde{Y}^\dagger\,\tilde{N}\,\dot{\tilde{Y}}-\tilde{Y}^\dagger \,\tilde{M} \,\tilde{Y}\right)\,,
\end{equation}
where matrices $\tilde{K}$, $\tilde{M}$ and $\tilde{N}$ are $3\times3$ time-dependent real matrices. 

The kinetic matrix $\tilde{K}$ can be diagonalized in the manner described in Eqs.\eqref{eq:basisrotate}-\eqref{eq:diagonalkinetic}, giving the eigenvalues
\begin{eqnarray}
\tilde\kappa_1 &=& \frac{3\,a^2}{\Mpl^4(6-\Omega)(-\dot{H})}\,\left[\frac{(-\dot{H})\,(\bar\alpha-1)(6-\Omega)^2}{\frac{12\,k^2}{a^2}(\bar{\alpha}-1)+\Omega\,(6-\Omega)\,(3\,H^2+\dot{H})} +\frac{{\cal Q}^2}{k^2/a^2}\right]^{-1}\,,
\nonumber\\
\tilde\kappa_2 &=& \frac{4\,\Omega\,(3\,H^2+\dot{H})k^2}{r^2\left[\frac{12\,k^2}{a^2}(\bar{\alpha}-1)+ \Omega(6-\Omega)(3\,H^2+\dot{H})\right]}
\left[
\bar{\alpha} + \frac{\Omega(r^2-\bar{\alpha})(6-\Omega)(3\,H^2+\dot{H})}{\frac{12\,k^2}{a^2}\,(r-1)^2}
\right]\,,
\nonumber\\ 
\tilde\kappa_3 &=& \Omega +
\frac{\Omega\,(6-\Omega)^2(r^2-\bar{\alpha})(3\,H^2+\dot{H})}{\frac{12\,k^2}{a^2}(r-1)^2}
\,\left[
\bar{\alpha}+\frac{\Omega(r^2-\bar{\alpha})(6-\Omega)(3\,H^2+\dot{H})}{\frac{12\,k^2}{a^2}\,(r-1)^2}
\right]^{-1}\,.
\label{eq:kineticeigcanonical}
\end{eqnarray}
We notice that the canonical transformation did not affect the two kinetic terms corresponding to the scalar graviton and quasidilaton perturbations, so we have $\tilde{\kappa}_2=\kappa_2$ and $\tilde{\kappa}_3=\kappa_3$. On the other hand, the first kinetic term has changed with respect to \eqref{eq:kineticeig}.

Let us discuss the stability conditions by considering the subhorizon limit of the system. Although the initial system was shown to be UV stable in Sec.\ref{sec:scalars}, we wish to make sure that the canonical transformation did not flip the sign of the first kinetic term in the UV. For large momenta, the kinetic eigenvalue is:
\begin{equation}
\tilde{\kappa}_1 = \frac{3\,a^2}{\Mpl^4(6-\Omega)(-\dot H)} + {\cal O}(k^{-2})\,,
\end{equation}
where we used that ${\cal Q}=k/a +{\cal O}(k^0)$ in this limit. This kinetic term is always positive for a matter fluid with equation of state $w >-1$. Expanding the other matrices for large momenta, one can also show that the sound speed for all three perturbations are still given by \eqref{eq:cs2}, so the transformation did not introduce any gradient instability either.

On the other hand, the qualitative result at intermediate momenta and in particular in the IR is now dramatically different compared to the case in Sec.\ref{sec:scalars}. If ${\cal Q}$ is non-vanishing in the limit $k\to0$, i.e. if the second term in Eq.\eqref{eq:Q2def} is non-zero, then it determines the sign of the kinetic term, 
\begin{equation}
\tilde{\kappa}_1 = \frac{3\,k^2}{\Mpl^4(6-\Omega)(-\dot H)\,\left(c_s^2 - \frac{\dot{P}}{\dot{\rho}}\right)} + {\cal O}(k^{4})\,.
\end{equation}
In this case, the no-ghost condition for the matter $c_s^2 > \dot{P}/\dot{\rho}$ can be written as:
\begin{equation}
\dot{\chi} \left[(1+c_s^2)P_{,\chi}-c_s^2P_{,X\chi}\dot\chi^2\right]>0\,.
\label{eq:matter-stability-general}
\end{equation}
It should be noted that if this IR stability condition is satisfied, then the kinetic term $\tilde{\kappa}_1$ is manifestly positive at any given momenta, including intermediate ones. 

Although our scalar field action is general, it is not meaningful to impose the condition \eqref{eq:matter-stability-general} to arbitrary matter sectors, whose action may inherently be unstable regardless of its coupling to the massive gravity. The stability conditions should be evaluated for the specific problem at hand.
As an example, we consider shift symmetric k-essence field with $P(X,\chi)=P(X)$. This is a relevant example as it can be used to model irrotational fluids. In this case, ${\cal Q} =k/a$ and both terms in the first of Eq.\eqref{eq:kineticeigcanonical} are manifestly positive for $0<\Omega<6$, $\bar\alpha>1$ and $-1<w<1$. Remarkably, the IR ghost which was present for any perfect fluid in the analysis of Sec.\ref{sec:scalars} is removed when we use the density perturbation as the variable.

As a second example, we consider a canonical scalar field with a potential $V$, i.e.
\begin{equation}
P(X,\chi) = X - V(\chi)\,.
\end{equation}
In this case, the no-ghost condition \eqref{eq:matter-stability-general} becomes:
\begin{equation}
\dot{\chi} V(\chi) = \partial_t V(\chi) <0\,.
\end{equation}
In other words, if the potential decreases with time, there is no ghost instability in the matter sector at any momenta. It is interesting to note that for this example, the analysis in Sec.\ref{sec:scalars} did not reveal any ghost instability for the field perturbation, as we found that $\kappa_1$ in Eq.~\eqref{eq:kineticeig} is manifestly positive as long as $c_s^2=1$. 

We stress that a canonical transformation does not change the nature of an instability or the physical observables in general. An IR ghost instability in one picture can correspond to a tachyonic instability in another. The crucial point is to determine whether the instability is safe or not, for relevant cases. 
We proved the stability of the system in the UV and thus any potential instability in the matter sector is bound to appear in the IR in the form of a tachyonic instability in terms of fluid perturbations. We argue that such an IR instability is harmless as it would be a manifestation of the Jeans instability.

\section{Canonical scalar field with a potential}
\label{sec:kination}
We now examine closely the case when the matter field has a canonical kinetic term and a potential. This is the scenario previously considered by Ref.\cite{Motohashi:2014una}, which corresponds to
\begin{equation}
P(X,\chi)=X-V(\chi)\,,
\label{eq:Canonicalscalar}
\end{equation}
and leading to energy density $\rho = X+V$ and sound speed $c_s^2=1$.
In this construction, the scalar field is not assumed to be an analogue of a fluid, so we will be using our results from Sec.\ref{sec:scalars}. 

For this example, we can use the background equations of motion \eqref{eq:BG-FIXEDPOINT} to replace $\dot{H}$ and $V(\chi)$ in terms of $H$ and $\dot{\chi}$, and rewrite the eigenvalues \eqref{eq:kineticeig} of the kinetic matrix as
\begin{eqnarray}
\kappa_1 &=& (6-\Omega)\,\left(\frac{\Omega(6-\Omega-\Xi^2)+4(\bar{\alpha}-1)\frac{k^2}{a^2H^2}}{\Omega(6-\Omega-\Xi^2)^2+4\,(\bar{\alpha}-1)(6-\Omega)\frac{k^2}{a^2H^2}}\right)\,,\nonumber\\
\kappa_2 &=& \frac{\Omega\,a^2H^2\,(6-\Omega-\Xi^2)}{(6-\Omega)r^2(r-1)^2}
\left(
\frac{\Omega(r^2-\bar{\alpha})(6-\Omega-\Xi^2)+4\,\bar{\alpha}(r-1)^2 \frac{k^2}{a^2H^2}}{\Omega(6-\Omega-\Xi^2)+4\,(\bar{\alpha}-1)\frac{k^2}{a^2H^2}}
\right)\,,\nonumber\\
\kappa_3 &=& 2\,\Omega\,\left(\frac{3\,(r^2-\bar{\alpha})(6-\Omega-\Xi^2)+2\,\bar{\alpha}(r-1)^2\frac{k^2}{a^2H^2}}{\Omega\,(r^2-\bar{\alpha})(6-\Omega-\Xi^2)+4\,\bar{\alpha}(r-1)^2\frac{k^2}{a^2H^2}}\right)\,,
\label{eq:kineticeigscalar}
\end{eqnarray}
where in the spirit of Ref.~\cite{Motohashi:2014una}, we defined the dimensionless quantity
\begin{equation}
\Xi \equiv \frac{\dot{\chi}}{\Mpl H}\,.
\end{equation}
The no-ghost conditions for this case are then simply:
\begin{equation}
0<\Omega<6\,,\qquad
1<\bar{\alpha}<r^2\,,\qquad
6-\Omega-\Xi^2 >0\,.
\label{eq:Huconditions}
\end{equation}
It can be verified that the latter condition in this set-up corresponds to \eqref{eq:cond3b} for $\Omega<6$. 

The main focus of Ref.\cite{Motohashi:2014una} is on the second of (\ref{eq:Huconditions}), which provides a time dependent range for the $\alpha_\sigma$ parameter. The concern is that even when one starts in a stable regime, the system can evolve into an instability. This is indeed a valid concern and Ref.\cite{Motohashi:2014una} actually provides a specific example where it is justified. This example is a special case of the model \eqref{eq:Canonicalscalar} where the potential vanishes, i.e. a kinetic energy dominated scalar field. This allows the background evolution to reduce to two equations:
\begin{equation}
3\,H^2+\dot{H} = \frac{6\,m^2\rho_m}{6-\Omega}\,,
\end{equation}
and 
\begin{equation}
r-1 = \frac{2\,\Omega\,\rho_m}{(6-\Omega)J\,\xi}\,,
\end{equation}
hence, $r$ is constant. Since during the kination dominated stage, the expansion rate redshifts as $a^{-3}$, thus $\bar\alpha \propto a^{-6}$. Therefore at early times, $\bar\alpha$ will grow very rapidly with respect to the constant $r$, eventually violating the stability condition $\bar{\alpha}<r^2$.
The kinetic term where this condition arises from is the second of \eqref{eq:kineticeigscalar}, which can be rewritten for the massless case as
\begin{equation}
\kappa_2 = \frac{2\,\Omega\,a^2m^2\rho_m}{(6-\Omega)r^2(r-1)^2}\left[\frac{\Omega(r^2-\bar\alpha)m^2\rho_m+2\,\bar\alpha(r-1)^2 k^2/a^2}{\Omega\,m^2\rho_m +2\,(\bar\alpha-1) k^2/a^2}\right]\,.
\end{equation}
We see that the problem arises when the numerator vanishes. Assuming that $\bar\alpha$ can actually exceed the value of $r^2$, then the modes with physical momentum
\begin{equation}
\frac{k}{a} > p_c \equiv \sqrt{\frac{\Omega\,m^2\rho_m(\bar\alpha-r^2)}{2\,(r-1)^2\bar\alpha}}\,,
\end{equation}
will have a positive kinetic term. As the ratio $(\bar\alpha-r^2)/\bar{\alpha}$ converges to unity at early times, the sooner the condition $\bar\alpha<r^2$ breaks, fewer modes will be ghost--like.

Despite the apparent severity of this problem, the specific example of kination dominated universe is non-representative. In this case, since $r$ is constant, going sufficiently early in time will result in $\bar\alpha$ coinciding with the constant $r^2$, making the kinetic term $\kappa_2$ cross zero. On the other hand, as one goes back in time the expansion rate $H$ will eventually grow as high as the effective field theory cutoff of the theory $\Lambda_3 = (m^2\Mpl)^{1/3}$.
Assuming $m\sim H_0$, the effective theory is valid up to $\Lambda_3 \sim 10^{-40}\Mpl \simeq 10^{-19} {\rm MeV}$. For a conservative scenario where thermalization occurs shortly after inflation, the expansion rate will reach $\Lambda_3$ already during the radiation dominated stage.\footnote{To put this cutoff scale into perspective, the QCD phase transition occurs at $H\sim 10^{-16}{\rm MeV}$ (assuming $g_*\sim 100$ and $T=150 {\rm MeV}$)} Since this time is well after any kination field can be relevant, the kinetic terms of all perturbations will be of ${\cal O}(1)$ and positive. Therefore, except for the IR ghost in the matter sector which is equivalent to the standard Jeans instability, no ghost will emerge within the regime of validity of the effective field theory.

Although we can avoid the specific problem with the kination field in a standard cosmological scenario, it is still not clear whether the evolution would drive the system beyond the stability bounds, even if $r$ is time dependent. On the other hand, the appearance of the ghost is qualitatively different than the IR ghost in the matter sector discussed at the end of Sec.\ref{sec:scalars}. In the case of a violation of $\bar\alpha<r^2$, the change of sign in $\kappa_2$ is a result of a vanishing numerator, meaning that  the corresponding degree of freedom becomes infinitely strongly coupled. Thus the strong coupling scale is zero at the critical momentum $k/a=p_c$. In other words, even if at low energies, a matter source starts to drive the system away from the stability, we cannot reliably talk about the regime where the ghost in Ref.\cite{Motohashi:2014una} would show up.

The situation has actually another aspect. The IR stability bound $\bar\alpha<r^2$ is also the condition \eqref{eq:alphabarup} that is required by the existence of the cosmological solution. If this condition is violated, we get an inconsistency leading to an imaginary $\dot{f}/n$. Then the question is, if one has an evolution (such as the kination domination) that forces the quantity $\dot{f}^2/n^2$ to cross zero, whether our ansatze would still hold and the equations would prevent this, or whether we would depart from the cosmological construction. Ref.~\cite{Heisenberg:2015voa} addressed this problem by pointing out that at early times, there is a departure from the fixed point solution, giving rise to a non zero $\dot{\xi}$, which may prevent the evolution to reach the $\dot{f}/n=0$ point. In fact, as $r$ can no longer be solved algebraically away from the fixed point, this may resolve the issue of a constant $r$ in the kination example as well.

However, understanding the behavior of the solutions away from the fixed point is challenging. The reason is that in the full set of equations, the evolution of $r$ is given by \eqref{eq:EQS}, after combining it with Eqs.~\eqref{eq:EQA} and \eqref{eq:eqstuck0}. Although one can evolve the equations numerically starting with initial conditions away from the fixed point, the coefficient of the $\dot{r}$ term vanishes on the fixed point and the numerical evolution would hit a singularity as the late time asymptotic solution is approached. In fact, this is the same property that allows us to determine $r$ algebraically at late times. Although Ref.~\cite{Kahniashvili:2014wua} considered the background evolution away from the fixed point, the authors actually considered a very restricted part of the evolution, where they set $\dot{f}/n$=constant, effectively reducing the St\"uckelberg constraint to the original quasidilaton one.
The equation that is solved in Ref.~\cite{Kahniashvili:2014wua} to evolve $r$ directly comes from the constancy of $\dot{f}/{n}$, thus the singularity at the fixed point is never observed. 

Another possibility is that the system may evolve towards a fixed point with $\dot{f}/N=0$ as the r.h.s. of \eqref{eq:eqstuck0} approaches zero. In this case the final configuration with $\dot{f}/N=0$ does not allow us to choose the unitary gauge for the time variable. Nonetheless the extended fiducial metric defined in \eqref{eq:extended-fiducial-FLRW} remains regular as far as $\dot{\sigma}/N\ne 0$. By assuming that $\dot{\xi}/N\to 0$ as $\dot{f}/N\to 0$, one obtains
\begin{equation}
 \frac{1}{\Mpl}\frac{\dot{\sigma}}{N} \to H\,,\quad
 r \to \frac{\sqrt{\alpha_{\sigma}}}{m}\frac{H}{\xi}\,,
\end{equation}
and thus the extended fiducial metric is indeed regular at late time.  By further assuming that $\dot{H}\to 0$, $\rho\to 0$ and $P\to 0$, one finds a new de Sitter fixed point specified by constant values of $H$ and $\xi$, which are determined by the following two equations derived from the background equations of motion. 
\begin{equation}
 3\left(1-\frac{\Omega}{6}\right) H^2 = m^2\rho_m\,, \quad
  m\left(\sqrt{\alpha_{\sigma}}H-m\xi\right)J  = \Omega H^2\,.
\end{equation}
Whether this new de Sitter fixed point solution is stable or not is an interesting problem for the future work.

The moral of this Section is as follows. The instability associated with the purely kinetic scalar field is irrelevant in realistic cosmology, although with other matter fields one should still be wary regarding the violation of the stability conditions, which also determine whether the background solutions exist. In order to understand this problem better, we need a detailed and complete numerical study, which is beyond the scope of this paper.

\section{Discussion}
\label{sec:discussion}

In this paper we studied the perturbative stability of the extended quasidilaton cosmology, in the presence of a matter sector that consists of a generic k-essence type scalar field. The analysis was restricted to the late time attractor solution where the graviton mass term acts like a cosmological constant and the quasidilaton field modifies the gravitational constant. We found that in the UV the solution is stable for a wide range of parameters. Conversely, in the IR two distinct types of ghosts can emerge, depending on the dominant fields in the matter sector.

When the scalar field behaves like a perfect fluid, we found that the scalar field perturbations become ghost--like in the IR. However, unlike the UV ghosts, this is not a signal of a catastrophic instability; we showed that this ghost can be removed by an appropriate choice of matter variable, namely the density perturbation. 
On the other hand, our analysis is not sufficient to verify that this IR ghost in the field perturbation becomes a classical tachyonic instability for the density perturbation. The main reason is that we have a system of three fields, all coupled to each other with coefficients that are time dependent. In the Lagrangian picture, it is generically impossible to decouple these to obtain the full frequency eigenvalues. Although this can be achievable in the Hamiltonian picture, it is technically involved and beyond the scope of the paper. In addition, to show that ghost modes can be avoided, we have also shown that the gradient part of the eigenfrequencies do not lead to any instabilities, which should be read as a proof of the stability of the system in the UV. Any potential instability in the matter sector is thus bound to appear in the IR, in the form of a tachyonic instability. We argue that such an IR instability is harmless as it would be a manifestation of the Jeans instability in the context of massive gravity. 

The second type of IR ghost emerges in the gravity sector. Although the two stability conditions corresponding to this sector have the same form as in the vacuum case \cite{DeFelice:2013tsa}, the conditions now become time dependent especially at early times due to the contribution from the background matter fields. In particular, we considered the example of Ref.~\cite{Motohashi:2014una}, where a purely kinetic energy (kination) scalar field can lead to a violation of one of the stability conditions, resulting in an IR ghost. We found that this specific scenario can be avoided in standard cosmology as any kination field would be subdominant when the energy of the universe surpasses the EFT scale. Moreover, in more general scenarios where the matter evolution drives the system away from stability, the model becomes infinitely strongly coupled and the IR instability does not appear within the regime of validity of the EFT. 

On the other hand, this type of evolution raises some questions about the background evolution. Even before such a strong coupling is reached, the evolution needs to break the existence requirement for background solutions in Eq.\eqref{eq:alphabarup}. Such an outcome can be avoided if the full system of equations prevents the evolution to ever reach this problematic point, e.g. by driving the solution away from the fixed point. Such a resolution was considered in Ref.~\cite{Heisenberg:2015voa}, although it is challenging to verify this scenario in a concrete set-up. In order to fully determine the fate of the background solutions, it is necessary to develop a detailed numerical analysis of the full equations of motion, away from the late time attractor. If this open problem can be addressed, the extended quasidilaton theory would be the simplest massive gravity theory which can accommodate a stable, realistic cosmology.

\acknowledgments
We thank Tina Kahniashvili for clarifying their calculations in Ref.~\cite{Kahniashvili:2014wua}.
AEG acknowledges support by STFC grant ST/L00044X/1. 
KK is supported by the UK Science and Technologies Facilities Council grants ST/K00090X/1 and ST/N000668/1 and the European Research Council through grant 646702 (CosTesGrav). The work of SM was supported by Japan Society for the Promotion of Science (JSPS) Grants-in-Aid for Scientific Research (KAKENHI) No. 24540256, and by World Premier International Research Center Initiative (WPI), MEXT, Japan.

\end{document}